\documentclass{pasa}%
\usepackage{natbib}

\def\actaa{Acta. Astronom.} %
\def\aj{AJ}%
\def\apj{ApJ}%
\def\apjl{ApJ}%
\def\apjs{ApJS}%
\def\apss{Ap\&SS}%
\def\aap{A\&A}%
\def\aaps{A\&AS}%
\def\mnras{MNRAS}%
\def\nat{Nature}%
\title[The Age of the Galactic Bulge]{The Controversial Star-Formation History and Helium Enrichment of the Milky Way Bulge}
\author[Nataf et al.]{David M. Nataf$^1$\thanks{david.nataf@anu.edu.au}\\
\affil{$^1$Research School of Astronomy and Astrophysics, Australian National University, Canberra, ACT 2611, Australia}}%
\jid{PASA}
\doi{10.1017/pas.\the\year.2015}
\jyear{\the\year}

\begin{document}%
\begin{abstract}
The stellar population of the Milky Way bulge is thoroughly studied, with a plethora of measurements from virtually the full suite of instruments available to astronomers. It is thus perhaps surprising that alongside well-established results lies some substantial uncertainty in its star-formation history. Cosmological models predict the bulge to host the Galaxy's oldest stars for [Fe/H]$\lesssim -1$, and this is demonstrated by RR Lyrae stars and globular cluster observations. There is consensus  that bulge stars with [Fe/H]$\lesssim0$ are older than $t \approx10$ Gyr. However, at super-solar metallicity, there is a substantial unresolved discrepancy. Data from spectroscopic measurements of the main-sequence turnoff and subgiant branch, the abundances of asymptotic giant branch stars, the period distribution of Mira variables, the chemistry and central-star masses of planetary nebulae, all suggest a substantial intermediate-age population ($t \approx 3$ Gyr). This is in conflict with predictions from cosmologically-motivated chemical evolution models and photometric studies of the main-sequence turnoff region, which both suggest virtually no stars younger than $t \approx 8$ Gyr. A possible resolution to this conflict is enhanced helium-enrichment, as this would shift nearly all of the age estimates in the direction of decreasing discrepancy. Enhanced helium-enrichment is also arguably suggested by measurements of the red giant branch bump and the R-parameter. 
\end{abstract}
\maketitle%
\section{INTRODUCTION }
\label{sec:intro}

The Galactic bulge, as discussed elsewhere in this review, is a significant component of the Milky Way galaxy, and one of our best laboratories for studies of extremely metal-poor stars, interstellar extinction, abundances at the metal-rich end, and globular clusters. Those are among the reasons why the bulge has been investigated by not only rigorous theoretical models, but also virtually the full suite of instrumentation and tools available to astronomers: photometrically, chemically, and kinematically,by both broad and less-precise surveys as well as targeted and highly-precise surveys. 

Given this library of theoretical predictions and series of constraints from diverse chemical evolution models, one might expect that the star-formation history of the Milky Way bulge\footnote{Throughout this work, we exclude from discussion the stellar population within $\sim$100 parsecs of the Galactic centre, where significant and ongoing star formation is well-documented \citep{2007AcA....57..173P,2012ApJ...751..132C}.} would be a solved or nearly-solved problem, as age is one of the stellar properties that are of greatest interest to astronomers, arguably the most interesting property. And indeed, for a while there was a consensus that the result of an old bulge ($t \gtrsim 10$ Gyr) had been firmly established, to quote several well-cited manuscripts:
\begin{quotation}
\noindent ``The CMD [Colour-Magnitude Diagram] of BaadeÕs Window field indicate a uniformly old age for stars in the Galactic bulge, thus helping to settle the question of the formation of galactic bulgesÓ  -- \citet{1994A&A...285L...5R}.
\end{quotation}
\begin{quotation}
\vspace{-0.40cm}  \noindent ``The bulge of our Galaxy formed at the same time and even faster than the inner Galactic halo." -- \citet {1999Ap&SS.265..311M}
\end{quotation}
\begin{quotation}
\vspace{-0.40cm}  \noindent ``The population with non-disk kinematics (which we conclude to be the bulge) has an old main-sequence turnoff point, similar to those found in old, metal-rich bulge globular clusters, Ó -- \citet{2002AJ....124.2054K}
 \end{quotation}
\begin{quotation}
\vspace{-0.40cm} \noindent ``The bulge age, which we found to be as large as that of Galactic globular clusters, or $\gtrsim 10$ Gyr. No trace is found for any younger stellar population.Ó  -- \citet{2003A&A...399..931Z}.
 \end{quotation}
 
As we will discuss in this review, all of the observational and theoretical arguments contributing to this prior consensus not only remain in place, but are now even more firmly established by vastly superior data, constraints, and analytical methodologies, consistently concluding that no more than a trace fraction of bulge stars are younger than $t \approx 10$ Gyr.  That bulge stars are not only old but the oldest in the Galaxy is a cosmologially-motivated prediction with both firm observational backing from RR Lyrae stars and Galactic globular clusters as well as being uncontested otherwise. 
 
 However, at the metal-rich end the picture becomes murkier. In addition to these impressive arguments, there are some comparably impressive arguments that a substantial fraction of bulge stars are of intermediate age, $t \approx 3$ Gyr. These arguments are independently robust, and are derived from high-precision spectroscopic observations of the main-sequence turnoff and subgiant branch, the period distribution of Mira variables, the abundances of asymptotic giant branch stars, and the abundances and central star masses of planetary nebulae. This is a severe discrepancy and the analyses are too sophisticated and multi-pronged for the solution to be embarassingly trivial.
 
Thus we are left with three possibilities: catastrophic failure of stellar evolution models at one or more stages of stellar evolution, or catastophic failure of observational and analytical methods being the first two.  An alternative suggested in this review and elsewhere, of helium-enrichment substantially higher than assumed by isochrones, would necessitate a catastrophic failure of chemical evolution models, likely in the assumed outputs of one or more of the sources of yields. 

The structure of this review is as follows. In Sections \ref{sec:ChemicalEvolution}-\ref{sec:Globulars} I present the arguments for an old bulge, whereas in Sections \ref{sec:microlens}-\ref{sec:AGB} I present the arguments for a substantial intermediate-age stellar population in the bulge. In Sections \ref{sec:RGBB}-\ref{sec:Rparameter} I present the independent evidence regarding helium-enrichment. The conclusion is presented in Section \ref{sec:Discussion}.

\section{Argument for an Old Bulge:  Chemical Evolution Models}
\label{sec:ChemicalEvolution}
Chemical evolution models have ubiquitously argued for a entirely old or nearly-entirely old old bulge \citep{1999Ap&SS.265..311M}, predominantly due to measurements of $\alpha$-element abundances \citep{1994ApJS...91..749M}, as well as dynamical arguments for inside-out Galaxy formation \citep{1962ApJ...136..748E,1978ApJ...225..357S,2010ApJ...708.1398T}, for which some of the detailed predictions are now being followed-up and investigated \citep{2014MNRAS.445.4241H}.

\citet{2011ApJ...729...16K} presented results of chemodynamical simulations of a Milky Way-type galaxy. The model adequately predicts the [O/Fe]-[Fe/H] relation for the bulge. The predicted bulge star-formation history peaks $\sim$12 Gyr ago. Star-formation continues though to the present day albeit at a slow pace, with a star-formation rate two orders-of-magnitude below the peak level. Some $\sim$50\% of bulge stars are at least 12 Gyr old,  $\sim$80\% of bulge stars are at least 10 Gyr old, and $\sim$95\% of bulge stars are at least 6 Gyr old. 

\citet{2012MNRAS.422.1902I} used an N-body/smoothed particle hydrodynamics code to study how a bulge could form in a "clumpy galaxy", whereby a gas-rich Galactic disk spontaneously forms massive clumps of stars \citep{1998Natur.392..253N}. It has been predicted that many of these clumps migrate tinward and coalesce into a bulge \citep{1999ApJ...514...77N}. These are predicted to be important: \citet{2014MNRAS.443.3675M} find that 83\% of the high-redshift star formation in their sample of 29  galaxies simulated with adaptive-mesh refinement takes place in compact, round clumps. Within this framework, stars 2 Kpc from the Galactic centre form (in the mean) within 2 Gyr of the start of the simulation of \citet{2012MNRAS.422.1902I}, and $\sim$85\% form within 2.75 Gyr. This predicted bulge also has an exponential density profile, a barred and boxy shape, nearly cylindrical rotation, albeit with a high metallicity, [M/H]$\approx +0.30$ in the mean. However, their simulation was stopped after 6 Gyr, once a smooth galaxy formed, and thus may not be sensitive to ``late" star formation. 

\begin{figure*}
\begin{center}
\includegraphics[totalheight=0.50\textheight]{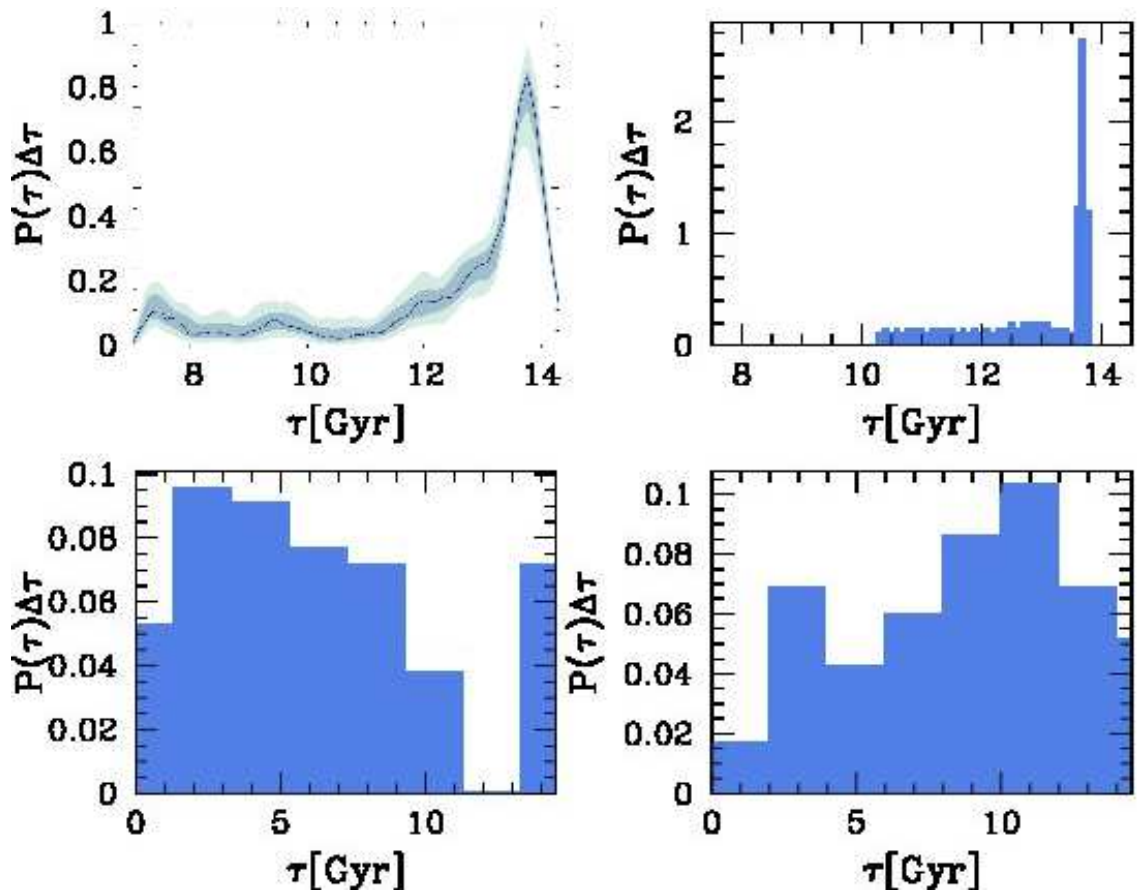}
\caption{Four distinct star-formation histories of the Galactic bulge as determined by four methodologies. \textbf{Top Left:}  Figure courtesy of Mario Gennaro, method described by \citet{2015arXiv150404865G}. The bulge star-formation history  derived from probabilistic modelling of five-band photometry extending deep below the main-sequence turnoff. The peak age is $t \sim 14$ Gyr, with a tail extending down to $t \sim 7.5$ Gyr . \textbf{Top Right:} Predicted bulge star-formation history from the chemical evolution model of \citet{2012A&A...548A..60G}. The peek age is $t \sim 13.5$ Gyr, with a tail extending down to $t \sim 10$ Gyr     \textbf{Bottom Right:} Age distribution of bulge stars from the sample of \citet{2013A&A...549A.147B}, no correction for selection effects is applied. The peak age is $t \sim 11$ Gyr, with a tail extending down to the present day. \textbf{Bottom Left} Bulge star-formation history from the planetary nebulae sample of \citet{2014A&A...566A..48G}, with full correction for selection effects. The peak age is  $t \sim 3$ Gyr, with a tail extending to older ages. }
\label{AgeHistMerged4}
\end{center}
\end{figure*}

The analytic model of \citet{2012A&A...548A..60G} is constrained by the observed abundances of Fe and Mg in the bulge \citep{2011A&A...534A..80H}. The prediction is that the metal-poor stars should have formed extremely rapidly, but the metal-rich component should continue forming stars over a longer timescale, with a spread in ages of ${\Delta}t \approx$ 4 Gyr. We show the predictions  of \citet{2012A&A...548A..60G} in the top-right panel of Figure 1. Similar conclusions are reached when comparisons are made to updated abundances of light, $\alpha$, and Fe-peak elements \citep{2013ApJ...765..157J,2014AJ....148...67J}. Further convincingly, measurements of a subsolar [La/Eu] ratio over the entire [Fe/H] range suggest a rapid formation timescale of less than 1 Gyr \citep{2012ApJ...749..175J}, as the measured ratios suggest a chemical evolution dominated by the r-process rather than the s-process. Moreover, as the ratio is virtually flat over the entire metallicity range, the relative contributions are expected to be the same as well, suggesting similar enrichment sources and thus a rapid chemical evolution. 

An exception to these consistent predictions of a nearly purely old bulge is that of \citet{2014ApJ...787L..19N}, who use an N-body+smoothed hydrodynamics code for their simulation, within which a bulge forms from the disk via dynamical instabilities. This models predicts that the bulge does include stars younger than 5 Gyr, and that they are disproportionately located closer to the plane. 

\section{Argument for an Old Bulge:  Photometry of the Main-Sequence Turnoff}
\label{sec:MSTO}
Photometric age estimates of the bulge first emerged with \textit{Hubble Space Telescope} (HST) photometry of Baade's Window by \citet{1993AJ....106.1826H}, who estimated a mean age of the bulge of 5 Gyr, but acknowledged possible errors in reddening, metallicity, and photometric zero points. \citet{2002AJ....124.2054K} followed-up with subsequent photometric epochs to measure proper motions, and thus kinematically separate the foreground disk stars from the background bulge stars. They concluded that the young stars identified by \citet{1993AJ....106.1826H} were disk contaminants, and that the bulge stellar population itself had an old main-sequence turnoff, comparable to that in globular clusters. \citet{2003A&A...399..931Z} followed-up with deeper HST and ground-based photometry, which included both optical and near-IR data, and reached a similar conclusion that the age of the bulge is nearly entirely composed of stars older than 10 Gyr. 

The data quality and analysis have been steadily been improving since, with further HST data obtained by \citet{2008ApJ...684.1110C} and \citet{2010ApJ...725L..19B}. \citet{2011ApJ...735...37C} analyzed the time-series photometry of the possible young stars that are more luminous than that of a 10 Gyr main-sequence turnoff. They found that the stellar variability in this population was characteristic of blue stragglers. They concluded no more than 3.4\% of Galactic bulge stars are younger than $\sim$5 Gyr, even under the most conservative estimates of the blue straggler normalization. \citet{2013A&A...559A..98V} analyzed deep ground-based photometry toward the corners of the boxy bulge (where reddening and crowding are less problematic), and where a potentially different chemodynamic subset of the Galactic bulge is being probed. They also found a stellar population consistent with a purely old age. 

Gennaro et al. (2015, in prep) is analyzing the five-band photometry  (F390W, F555W, F814W, F110W, F160W) first presented by \citet{2010ApJ...725L..19B}, using a hirearchical Bayesian formalism described in \citet{2015arXiv150404865G}. Preliminary results, which incorporate constraints from the spectroscopic metallicity-distribution-function of the Galactic bulge and its measured [$\alpha$/Fe]-[Fe/H] trends, are suggestive of a nearly purely old bulge. The star-formation history is shown in the top-left panel of Figure 1, it peaks very early after the big bang with a tail extending down to $\sim$7 Gyr. Roughly $\sim$50\% of the stars are older than $\sim$12.5 Gyr, and $\sim$80\% are older than 10 Gyr. No stars are younger than $\sim$7 Gyr. 

These constraints are mpressive -- and they are spectacularly inconsistent with age determinations from other means which we will discuss in subsequent sections. Motivated by this discrepancy and by the work of \citet{2010ApJ...714.1072M}, \citet{2012ApJ...751L..39N} suggested that enhanced helium-enrichment could be a cause of the discrepancy. It was shown that analyzing stellar populations with isochrones that assumed a lower fractional helium abundance $Y$ than the actual value would yield overestimated photometric ages and underestimated spectroscopic ages, with the photometric age error going as $t_{\rm{Inferred}}/t_{\rm{True}} \approx 1 + 2.8{\Delta}Y$, where ${\Delta}Y$ is the difference between the actual initial helium abundance of the stellar population and that of the isochrones used to study the population.

% \citet{2015A&A...577A..72V} subsequently showed that the offset to spectroscopic ages would be smaller than $\sim$1 Gyr in the mean. 

\section{Argument for an Exceptionally Older Bulge at Low Metallicities: Horizontal Branch Morphology}
\label{sec:HBmorphology}
\citet{1992AJ....104.1780L} showed that the RR Lyrae stars observed toward Baade's window had metallicities that were $\sim$0.6 dex higher than at Galactocentric distances of $8\,\rm{Kpc} \leq R_{GC} \leq 40\,\rm{Kpc}$. Stellar evolution models were run to show that this predicted a mean age difference of ${\Delta}t = 1.3 \pm 0.3$ Gyr.  This offset is entirely one in relative ages, and is as such less insensitive to issues such as distance, reddening, mass-loss, primordial helium abundance, and rotation. 

The higher metallicities of Galactic bulge RR Lyrae have since been confirmed by \citet{2008AJ....136.2441K}, \citet{2012ApJ...750..169P}, and \citet{2014arXiv1412.4121P} using substantially larger databases. In a preliminary analysis of $\sim$100 bulge RR Lyrae, \citet{2015ApJ...808L..12K} found that the RR Lyrae with the most peculiar dynamics including a halo-like orbit, also had one of the lowest metallicities in the sample -- $\sim$0.60 dex below the mean of [Fe/H]$\approx -1.00$. 

At the metal-rich end, the horizontal branch is a less effective tool to constrain the age-metallicity relation of a stellar population, since the colour and magnitude of the red clump nearly level off with shifts in age and metallicity (\citealt{2014MNRAS.442.2075N}, results derived from \citealt{2004ApJ...612..168P} and \citealt{2006ApJ...642..797P}). 

\section{Argument for an Older Bulge for [Fe/H]$\lesssim 0$: Tracing Relative Ages with Globular Clusters}
\label{sec:Globulars}
It is conceivable that ages of globular clusters might yield a coarse estimate of the star-formation history of field stars, though the details of the correspondence are not yet worked out. The advantage of globular clusters is that they are relatively simple stellar populations, located at a fixed distance, thus reducing the range of free parameters needed to interpret them in a stellar population framework. 

\citet{2009A&A...507..405B} analyzed VLT-Flames spectroscopy of 8 red giants in the bulge globular cluster NGC 6522, and derived abundances of [Fe/H]$=-1.0 \pm 0.20$ and [$\alpha$/Fe]$\approx +0.25$, since confirmed by \citet{2014MNRAS.445.2994N} who measured [Fe/H]$=-1.15\pm0.16$. This gives the cluster an exceptionally high metallicity for one with a blue horizontal branch morphology, suggestive of a substantially older age than the mean Galactic globular cluster age-metallicity relation.  \citet{2006A&A...449..349B} and \citet{2007AJ....134.1613B} presented nearly identical arguments for the Galactic bulge globular clusters HP-1 and NGC 6558, and thus the indication of an old relative age at [Fe/H]$=-1.0$ is a trend and not an anomaly.  Interested readers are referred to \citet{2010ApJ...708..698D} and \citet{2014ApJ...785...21M} for reviews for the link between age and horizontal branch morphology in Galactic globular clusters.

At higher metallicities, there is the case of NGC 6528. \citet{2014ApJ...782...50L} estimated $t = 11 \pm 1$ Gyr and [Fe/H]$\approx +0.20$ from deep HST photometry of the globular cluster. Their analysis was supported by several diagnostics such as the red clump, the red giant branch bump, the shape of the subgiant branch and the slope of the main-sequence, which are minimally affected by uncertainties in reddening and distance. \citet{2014A&A...565A...8C} estimated [Fe/H]$=-0.04$ for the cluster using stromgren photometry, following which they estimate an age of $t=10-12$ Gyr. 

Another interesting globular cluster is that of Terzan 5, for which the metallicity-distribution function \citep{2014ApJ...795...22M} is unnervingly similar to that of the Milky Way bulge as a whole  \citep{2013MNRAS.430..836N}. The two dominant populations have metallicities of [Fe/H]$\approx -0.30$ and [Fe/H]$\approx +0.25$, with the more metal-rich horizontal branch being $\sim$0.30 mag brighter in $K$-band. This luminosity offset can either be explained by an age offset, such that the metal-rich population is $\sim$6 Gyr old \citep{2009Natur.462..483F}, or enhanced helium-enrichment, such that the metal-rich population has $Y \sim 0.33$ \citep{2010ApJ...715L..63D}. %This dichotomy mirrors that which we find elsewhere in the bulge, and has at this time not been solved. 

\section{Argument for an Intermediate-age Bulge at Super-solar Metallicities: Spectroscopic Measurements of Microlensed Turnoff and Subgiant Stars}
\label{sec:microlens}
Main-sequence turnoff and subgiant branch stars in the Galactic bulge have a typical apparent magnitude of $V \sim 20$, and are thus prohibitively expensive targets for investigation with high-resolution spectroscopy. That is why isochrone-based age estimated from atmospheric parameters, which are most accurate and most precise for stars on the main-sequence turnoff and subgiant branch \citep{2013MNRAS.429.3645S,2014A&A...565A..89B,2014arXiv1412.3453M,2015arXiv150407992S}, have until recently not been completed for Galactic bulge stars. 

Large microlensing surveys have changed this landscape, since every year thousands of Galactic bulge stars are observed to be brightened by factors of up to $\sim2000 \times$ \citep{2013ApJ...778..150S,2015AcA....65....1U}. There are now published atmospheric parameters from high-resolution, high signal-to-noise spectroscopy for 58 MSTO+SGB stars toward the Galactic bulge \citep{2013A&A...549A.147B}. The typical measurement errors on the atmospheric parameters are 0.1 dex in [Fe/H], 0.20 dex in $\log{g}$, and 100 Kelvin in $T_{\rm{eff}}$.  This combines for typical age measurement errors of 3 Gyr, where the error is from that in the atmospheric parameters and does not include systematic errors from other effects. 

\citet{2013A&A...549A.147B} argue that ``metal-poor" Galactic bulge stars, which they define as those with [Fe/H] $\leq -0.10$, are nearly all old, with ages of $10 \lesssim t/\rm{Gyr} \lesssim 12$. More metal-rich stars are argued to show a larger distribution, with ages of $2 \lesssim t/\rm{Gyr} \lesssim 10$, with a peak around $t = 4-5$ Gyr. Young stars are detected along both the turnoff and the subgiant branch. We show the age distribution measured by \citet{2013A&A...549A.147B} in the bottom-right panel of Figure 1.

At this time the results from these microlensed source stars have survived several challenges, \citet{2010ApJ...711L..48C} first observed a surprising correlation between the peak magnification of microlensing events and the metallicity of the source star, suggesting some undiagnosed systematic. The correlation is now weaker than it originally was, and partly explained by systematics \citep{2013A&A...549A.147B}. The combined effects of limb darkening and differential magnification across the source star surface have been shown to have a negligible impact on the derived atmospheric parameters of the source star, for example the error on [Fe/H] is no greater than 0.03 dex \citep{2010ApJ...713..713J}. \citet{2012ApJ...751L..39N} suggested that enhanced helium-enrichment could be biasing the age determinations, but \citet{2015A&A...577A..72V} showed that the mean offsets are expected to be smaller than $\Delta t = 1$ Gyr, even for ${\Delta}Y/{\Delta}Z=5$. \citet{2015A&A...577A..72V} also show that the expected offsets due to changes in the mixing length or diffusion are small. Finally, \citet{2015A&A...577A..72V} have independently estimated the best-fit ages of the \citet{2013A&A...549A.147B} sample, and also found a substantial young population, with 16 of the 58 stars having best-fit ages $t \leq 5$ Gyr. 

\section{Arguments for an Intermediate-Age Bulge:  Asymptotic Giant Branch Stars and Planetary Nebulae}
\label{sec:AGB}
Asymptotic giant branch and planetary nebulae (PNe) are a distinct window into the star formation history of a stellar population. 

\citet{1991MNRAS.248..276W} used four years of near-IR photometry to study the pulsational periods of Mira variables observed toward the Galactic bulge. Periods were determined for 104 objects, of which the mode of the period distribution was $\sim$475 days, with a minimum and maximum of 170 and 722 days respectively. This was contrasted to the periods of $\sim$250 days associated with old (thick disk, halo, globular cluster) populations, and the much longer periods (up to 2000 days) associated with young stellar populations in the thin disk \citep{1987ASSL..132...33F}. They estimated a mean age for the progenitor of the bulge Mira population of $t \sim 3$ Gyr, and brought up initial metallicity and helium abundance as sources of substantial uncertainty. More recently, \citet{2009MNRAS.399.1709M} analyzed a sample 1,364 Mira variables toward the Galactic centre, so much closer to the plane. Their period distribution spans the range $100 \lesssim P/\rm{days} \lesssim 630$, with the mode to their distribution of $P \sim 300$ days, suggestive of even younger ages than the work of \citet{1991MNRAS.248..276W}.

\citet{2014A&A...566A..48G} inferred the white dwarf  / central star of the planetary nebulae masses of 31 Galactic bulge PNe, from which initial stellar masses (and thus coarse ages) were estimated by means of an initial-final mass relations, Upon using an empirical initial-final mass relation from clusters \citep{2009MNRAS.395.1795C}, \citet{2014A&A...566A..48G} derived a star-formation history for the progenitor population of the bulge PNe that they deem implausible. Even after correcting for selection effects such as the mass-dependent lifetime of the PNe phase, they inferred age distribution is sharply peaked at $t \sim 3$ Gyr. \citet{2014A&A...566A..48G} applied the correction of shifting the initial-final mass relation of bulge stars to agree at the end points with the age measurements of \citet{2013A&A...549A.147B}. They found that the star-formation history of the Galactic bulge peaked $\sim$3 Gyr ago, with a very slow decline in number counts to ages of $\sim$11 Gyr, and then a peak at 15 Gyr. In their conclusion, \citet{2014A&A...566A..48G} state that the metal-poor component of the bulge is underrepresented among bulge PNe.

A recent, similar investigation is that of \citet{2013MNRAS.428.2577B}, which is a meta-analysis of previously-published bulge PNe data \citep{2000A&A...353..543C,2004A&A...414..211E,2004MNRAS.349.1291E,1992A&A...255..255R,1997A&AS..126..297R,2001MNRAS.327..141L,2007MNRAS.381..669W} and a set of updated models developed in that work. The method of \citet{2013MNRAS.428.2577B} accounts for both the inferred white dwarf mass and the observed abundances (including) helium of the PNe. \citet{2013MNRAS.428.2577B} finds that the best-fit models to reproduce the mean of the distribution have main-sequence progenitor masses of $M \sim 1.5 M_{\odot}$ and initial helium abundances of $Y \sim 0.32$, suggesting a peak age of $t \sim 3$ Gyr and a helium-to-metals enrichment ratio $dY/dZ \approx 4$. These models also match the asymptotic giant branch tip luminosity of the Galactic bulge. 

A third investigation of the kind is that of \citet{2015MNRAS.449.1797D} , who studied 20 PNe in the bulge and disk and also measured abundances of Ar, C, Cl, He, N, Ne, and O relative to H. Their best-fit models suggest that their bulge sub-sample, with helium abundances lying in the range $0.28 \lesssim Y \lesssim 0.43$, are best-fit by progenitors with initial masses $M \sim 2-4 M_{\odot}$. They consider these initial masses unlikely given other constraints on the star-formation history of the bulge, and point to the analysis of  \citet{2013MNRAS.428.2577B} who used He-enhanced models to match the bulge population of PNe.

A different diagnostic was explored by \citet{2007A&A...463..251U}: spectroscopic abundances of asymptotic giant branch stars. Technetium, an unstable element synthesised via the s-process, was identified in the spectra in 4 of the 27 stars studied. \citet{2007A&A...463..251U} stated that this required third dredge-up, and thus relatively large initial masses ($M \gtrsim 1.5 M_{\odot}$) and young ages for a sizable fraction of the bulge stellar population. Finally, \citet{2015A&A...579A..76J}  has recently fit evolutionary and dust models to multi-wavelength observations of the reddest Galactic bulge asymptotic giant branch stars, and estimated an initial mass range of $1.1 \lesssim M/M_{\odot} \lesssim 6.0$.

A concern with these claims is the sensitivity of stellar model predictions to chemistry. \citet{2014ApJ...784...32K} showed that the theoretical initial-final mass relation is very sensitive to initial helium abundance.  At fixed initial mass, helium-enriched stars have shorter lifetimes and leave behind more massive white dwarfs. \citet{2014MNRAS.445..347K} subsequently showed that carbon-star prediction and third dredge-up both become less likely as initial metallicity and/or initial helium abundance are increased. 

Regardless of the largely unexplored uncertainties, diagnostics from Galactic bulge asymptotic giant branch stars consistently argue for a stellar population with substantial and possibly predominant star formation between 2 and 8 Gyr ago. 

\section{Argument for a Helium-Enhanced Bulge: The Red Giant Branch Bump}
\label{sec:RGBB}
The red giant branch bump of a stellar population occurs during the first ascent of the red giant branch, resulting from the hydrogen-burning shell moving through the composition discontinuity left by the deepest penetration of the convective envelope \citep{1997MNRAS.285..593C,2014MNRAS.445.3839N,2015ApJ...804....6G,2015MNRAS.450.2423A}. 

The red giant branch bump of the Galactic bulge stellar population was first measured by \citet{2011ApJ...730..118N}, a detection since confirmed \citep{2011ApJ...735...37C,2011A&A...534L..14G,2013MNRAS.435.1874W}. This was done as part of an attempt to clean up the colour-magnitude diagram so as to better model the spatial morphology of the bulge, but a surprise emerged: the red giant branch bump had anomalous properties. The number counts were very low, and the luminosity a little bright. The more thorough and more detailed investigation of \citet{2013ApJ...766...77N} followed, which focused on an  investigation of the bump properties in 72 Galactic globular clusters to use as an empirical calibration. \citet{2013ApJ...766...77N} found that the red giant branch bump of the Galactic bulge is 0.74 mag fainter than the red clump whereas the expectation was 0.84 mag fainter, and had number counts 20\% that of the red clump whereas the expectation was 26\%. 

Stellar models \citep{2012ApJ...746...16V} were run to ascertain whether a mean age or mean helium offset could better match the discrepancy with globular clusters. Shifting either the mean helium by ${\Delta}Y=+0.06$ or the mean age by ${\Delta}t=-5$ Gyr would solve the discrepancy in number counts, but whereas the helium shift also solved the discrepancy in luminosity, the age shift would brighten the bump by $\sim$0.30 mag where the offset is only $\sim$0.10 mag. 

Thus, the bulge red giant branch bump is more consistent with enhanced helium than a younger age. With that said, there is substantial uncertainty from factors such as correlations between younger age and enhanced helium, extrapolation of the empirically-calibrated bump properties to higher metallicity, and measurement uncertainty of the bulge metallicity distribution function at the high-metallicity end. 

\section{The R-parameter and the Helium Abundance of the Galactic Bulge}
\label{sec:Rparameter}
The comparison between the number of red giant branch stars and horizontal branch stars yields a constraint on the helium abundance of a stellar population, as it is a steeply-sensitive function of helium abundance and largely insensitive to age \citep{1994A&A...285L...5R}. The metallicity contribution can be factored out, as metallicity is measurable by other means. 

The first estimate is that of \citet{1994A&A...285L...5R}, who estimated $Y=0.325 \pm 0.025$, where the metal-poor globular cluster M3 (NGC 5272) was used as the zero-point. \citet{1995A&A...300..109M} estimated $Y=0.28 \pm 0.02$, by analyzing near-IR photometry from five bulge fields, and concluded that he helium abundance of the bulge is comparable to that of the metal-rich clusters 47 Tuc (NGC 104) and M69 (NGC 6637). Simultaneously and independently, \citet{1995AJ....110.2788T} estimated $Y=0.27 \pm 0.03$, by comparing their $K$-band luminosity functions to theoretical models.  

%. However, the zero-point was calibrated by assuming that the globular cluster M3 (NGC 5272) had an initial helium-abundance of Y$=0.23$ and a metallicity 2 dex lower than that of the bulge. Updating the estimates to ${\Delta}\rm{[M/H]}=1.50$ dex \citep{2005AJ....129..303C} and $Y=0.25$ would shift the inferred helium abundance higher. 

Given two decades of substantial progress in virtually all of the sources of uncertainty affecting these determinations, it would be interesting to see the estimates updated. 

\section{Discussion and Conclusion}
\label{sec:Discussion}

\begin{figure*}
\begin{center}
\includegraphics[totalheight=0.57\textheight]{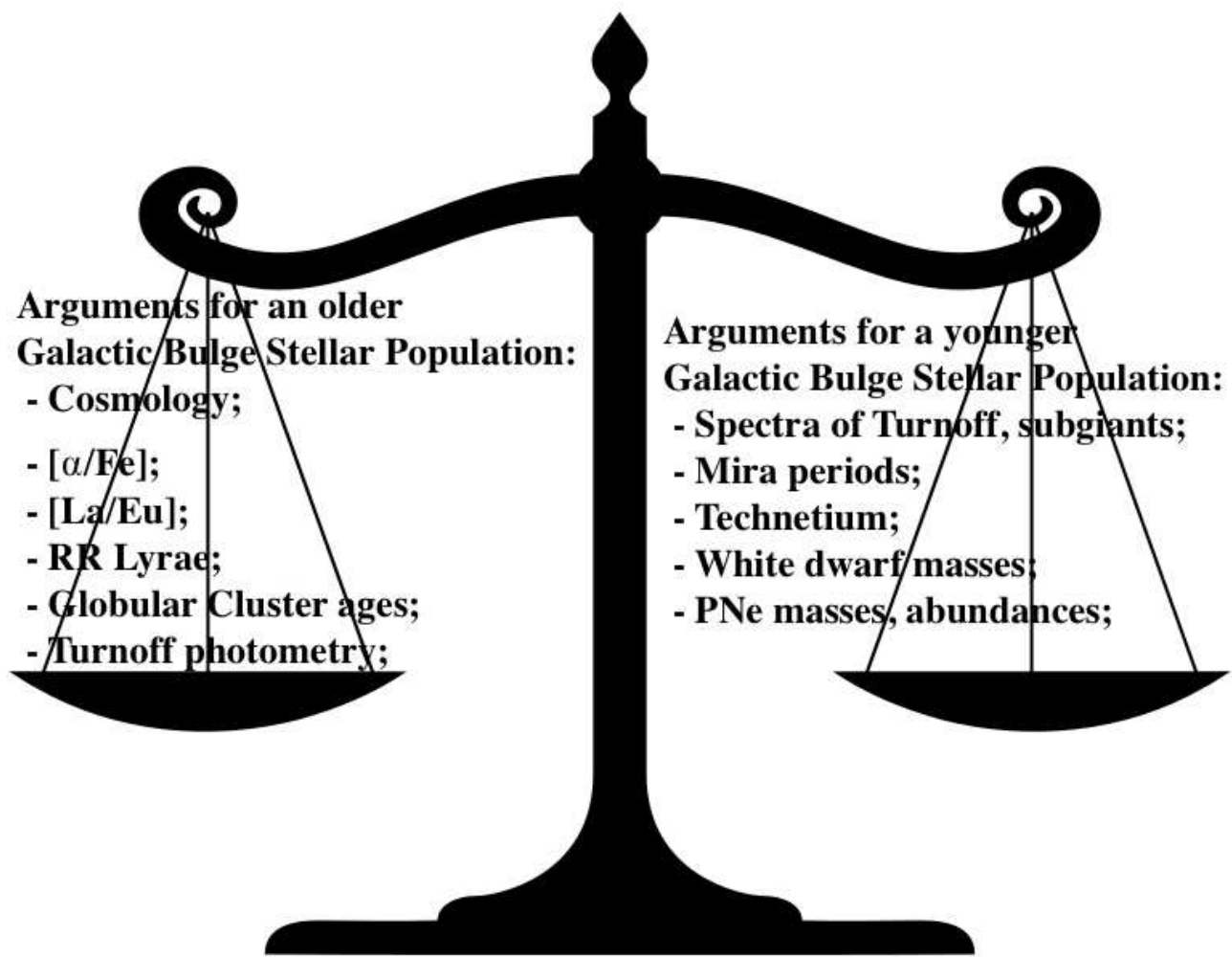}
\caption{On the left-hand side of the scale are summarized the arguments for an old bulge, that are pertinent to some or all of the bulge stellar population. On the rght-hand side are summarized the arguments for a substantial intermediate-age component to the bulge stellar population.}
\label{Scales}
\end{center}
\end{figure*}

The theoretical and observational arguments as to the age of the bulge have been presented. The state of consensus or disagreement is largely a function of metallicity. At low metallicities, there is a consensus from both theory and observations (e.g. RR Lyrae metallicities) that the oldest stars in the Galaxy are to be predominantly found in the bulge, with rapid enrichment of the gas from which stars formed and thus star formation up to [Fe/H]$=0$. In contrast there is controversy at super-solar metallicities, with various lines of evidence suggesting nearly no stars younger than $t=8$ Gyr, and other lines of evidence suggesting a substantial intermediate-age ($t \approx 3$ Gyr) population. The source of controversy is summarized in Figure 2. Enhanced helium enrichment has been suggested due to observations of the main-sequence turnoff morphology, the red giant branch, the R-parameter, and the central star masses of planetary nebulae, though more analysis is needed. 

Solutions to these discrepancies could emerge from new observational probes \citep{2012AcA....62...33N}, and/or a more comprehensive and rigorous suite of stellar evolution models \citep{2015A&A...577A..72V}. 

\section*{Acknowledgments}
DMN was supported by the Australian Research Council grant FL110100012.

\end{document}